\documentstyle[multicol,prl,aps,epsfig]{revtex}
\begin{document}
\draft

\title{An Electronic Mach-Zehnder Interferometer }
\author{Yang Ji, Yunchul Chung, D. Sprinzak,
M. Heiblum, D. Mahalu, and Hadas Shtrikman }
\address{Braun Center for Submicron Research, Department of Condensed
 Matter Physics, Weizmann Institute of Science, Rehovot 76100, Israel}
\date{\today}
\maketitle

\begin{abstract}
\textit{Double-slit} electron interferometers, fabricated in high
mobility two-dimensional electron gas (2DEG), proved to be very
powerful tools in studying coherent wave-like phenomena in
mesoscopic systems~\cite{r1,r2,r3,r4,r5,r6}. However, they suffer
from small fringe visibility due to the many channels in each slit
and poor sensitivity to small currents due to their \textit{open
geometry}~\cite{r3,r4,r5,r7}. Moreover, the interferometers do not
function in a high magnetic field, namely, in the quantum Hall
effect (QHE) regime~\cite{r8}, since it destroys the symmetry
between left and right slits.  Here, we report on the fabrication
and operation of a novel, single channel, two-path electron
interferometer that functions in a high magnetic field. It is the
first \textit{electronic analog} of the well-known optical
\textit{Mach-Zehnder} (MZ) interferometer~\cite{r9}.  Based on
single \textit{edge state} and \textit{closed geometry} transport
in the QHE regime the interferometer is highly sensitive and
exhibits very high visibility (62\%). However, the interference
pattern decays precipitously with increasing electron temperature
or energy. While we do not understand the reason for the dephasing
we show, via shot noise measurement, that it is not a decoherence
process that results from inelastic scattering events.

\end{abstract}

\begin{multicols}{2}

Direct phase measurements of electrons, customarily done in
double-slit interferometers~\cite{r1,r2,r3,r4}, are difficult to
perform under strong magnetic fields.  Electrons are being
diverted by the Lorentz force, perform chiral skipping orbits, and
prefer one slit to the other - thus breaking the symmetry of the
interferometer. At the extreme quantum limit, namely, in the QHE
regime, the \textit{skipping orbits} quantize to
quasi-one-dimensional like states, named \textit{chiral edge
states}.  We exploited the chiral motion of the electrons and
constructed an electronic analog of the ubiquitous optical
Mach-Zehnder (MZ) interferometer~\cite{r9} (described
schematically in Fig. 1a).  A beam splitter \textbf{BS1} splits an
incoming monochromatic light beam from source \textbf{S} into two
beams, which, after reflection by mirrors \textbf{M1} and
\textbf{M2}, recombine and interfere at \textbf{BS2} to result in
two outgoing beams (collected by detectors \textbf{D1} and
\textbf{D2}). When the phase along one of the paths varies both
signals in \textbf{D1} and in \textbf{D2} oscillate out of phase,
and since no photons are being lost the sum of both signals stays
always equal to the input in \textbf{S}. In the electronic
counterpart, depicted in Fig. 1b, \textit{quantum point contacts}
(QPC) function as beam splitters and \textit{Ohmic contacts} serve
as detectors. A QPC is formed in the 2DEG by depositing a split
metallic gate on the surface of the semiconductor and biasing it
negatively with respect to the 2DEG. The induced potential in the
2DEG creates a barrier under the gate bringing the two oppositely
propagating edge currents to the small opening in the barrier,
allowing thus backscattering. As shown schematically in Fig. 1b
QPC1 splits the incoming edge current from \textbf{S} to two
paths, a transmitted \textit{inner path} and a reflected
\textit{outer path}, both later recombine and interfere in QPC2,
to result with two edge currents (collected by \textbf{D1} and
\textbf{D2}).

The actual device, seen in Fig. 1c, was fabricated in a high
mobility 2DEG embedded in a GaAs-AlGaAs heterojunction.  A
ring-shaped mesa, 3$\mu$m in width, was defined by plasma etching
with Ohmic contacts (for \textbf{S}, \textbf{D1}, and \textbf{D2})
connected to the inner and outer edges of the ring.  The inner
contact, \textbf{D2}, and the two QPCs are connected to outside
sources via air bridges that float above the mesa.  A phase
difference $\varphi$ between the two paths is introduced via the
Aharonov-Bohm (AB) effect~\cite{r10,r11}, $\varphi=2\pi
BA/\phi_0$, with $B$ the magnetic field, $A$ the area enclosed by
the two paths ($\sim 45\mu$m$^2$), and $\phi_0 =4.14 \times
10^{-15}$Tm$^2$ the flux quantum. A few \textit{modulation gates},
\textbf{MG}, are added above the outer path in order to tune the
phase $\varphi$ by changing the area $A$. We briefly review the
operation of the interferometer.  At filling factor 1 in the QHE
regime a single chiral edge state carries the current.  The
interfering current, in turn, is proportional to the transmission
probability from source to drain $T_{SD}$. Neglecting dephasing
processes and having the transmission (reflection) amplitude $t_i$
($r_i$) of the i$^{th}$ QPC fulfilling $|r_i|^2+|t_i|^2=1$,
then~\cite{r7} $I_{D1} \propto T_{SD1}= |t_1 t_2+r_1 r_2 e^{i
\varphi}|^2= |t_1 t_2|^2+|r_1 r_2|^2+2|t_1 t_2 r_1 r_2|cos\varphi$
and $I_{D2}  \propto T_{SD2} =|t_1 r_2+r_1 t_2
e^{i\varphi}|^2=|t_1 r_2|^2+|r_1 t_2|^2-2|t_1 t_2 r_1
r_2|cos\varphi$. Note that ideally the two currents oscillate out
of phase as function of $\varphi$ while $T_{SD1}+T_{SD2}=1$.  The
visibility of the oscillation is defined as:
$\upsilon=(I_{max}-I_{min})/(I_{max}+I_{min})$ and, for example,
when QPC2 is tuned so that $T_2=0.5$, the visibility is
$\upsilon=2\sqrt{T_1(1-T_1)}$.

Measurements were done at filling factor 1 (magnetic field
$\sim$5.5T) and also at filling factor 2 with similar results.
With a refrigerator temperature $\sim$6mK the electron temperature
was determined by measuring the equilibrium noise~\cite{r12} to be
$\sim$20mK.  High sensitivity measurement of the interference
pattern was conducted at $\sim$1.4MHz with a spectrum analyzer.
Current at \textbf{D1} (or \textbf{D2}) was filtered and amplified
in situ by LC circuit and a low noise home-made pre-amplifier,
both placed near the sample and cooled to 1.5K.\linebreak

\begin{figure}
\begin{center}
\leavevmode \epsfxsize=8.5 cm  \epsfbox{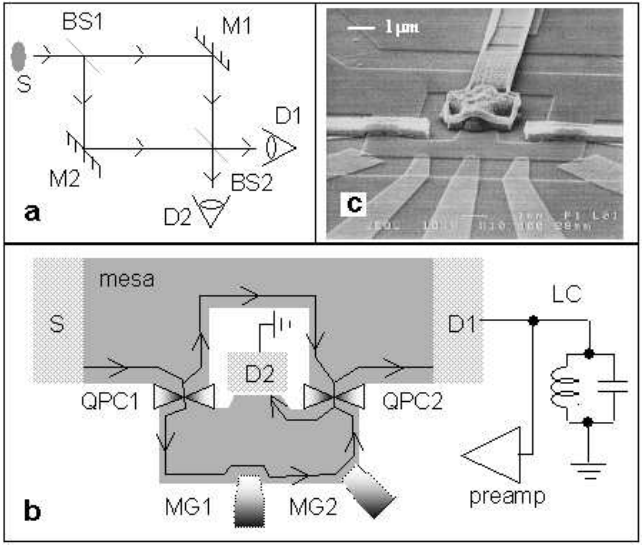}
\end{center}
\vspace{0 cm} \caption{The configuration and operation of an
optical Mach-Zehnder interferometer and its actual realization
with electrons.  (a) Schematics of an optical Mach-Zehnder
interferometer.  \textbf{D1} and \textbf{D2} are detectors,
\textbf{BS1} and \textbf{BS2} are beam splitters, and \textbf{M1}
and \textbf{M2} are mirrors. With 0($\pi$) phase difference
between the two paths, \textbf{D1} measures maximum (zero) signal
and \textbf{D2} zero (maximum) signal.  The sum of the signals in
both detectors is constant and equals to the input signal. (b)
Schematics of the electronic Mach-Zehnder interferometer and the
measurement system. Edge states are formed in a high perpendicular
magnetic field. The incoming edge state from \textbf{S} is split
by \textbf{QPC1} (quantum point contact) to two paths, of which
one moves along the inner edge and the other along the outer edge
of the device.  The two paths meet again at \textbf{QPC2},
interfere, and result in two complementary currents in \textbf{D1}
and in \textbf{D2}.  By changing the contours of the outer edge
state and thus the enclosed area between the two paths, the
modulation gates (\textbf{MG}) tune the phase difference between
the two paths via the Aharonov-Bohm effect.  A high
signal-to-noise-ratio measurement of the current in \textbf{D1} is
performed at 1.4MHz with a cold LC resonant circuit as a band pass
filter followed by a cold, low noise, preamplifier. (c) SEM
picture of the device.  A centrally located small Ohmic contact
($3 \times 3\mu m^2$), serving as \textbf{D2}, is connected to the
outside circuit by a long metallic air-bridge. Two smaller
metallic air-bridges bring the voltage to the inner gates of
\textbf{QPC1} and \textbf{QPC2} - both serve as beam splitters for
edge states. The five metallic gates (at the lower part of the
figure) are modulation gates (\textbf{MG}).}
\end{figure}

\noindent A standard lock-in technique, with a low-frequency
signal (7Hz, 10$\mu$V RMS), gave similar results, however, the
measurement lasted much longer and was prone to sample's
instability.  Since at 5.5T each flux quantum occupies and area of
some $10^{-15}m^2$ (some 60,000 flux quanta thread the area
\textbf{A}), a minute fluctuation in the superconducting magnet's
current or in the area would smear the interference signal.  Two
measurement methods were employed.  The first relied on the
unavoidable decay of the short-circuited current that circulates
in the superconducting magnet (being in the so called,
\textit{persistent current mode}).  In this mode the magnetic
field decays smoothly at a rate of $\sim$0.12mT/hour ($\sim$1 flux
quantum every 50 minutes).  The second was via scanning the
voltage on a modulation gate at a rate much faster than the decay
rate of the magnetic field, thus changing the area $A$, the
enclosed flux, and consequently the AB phase.\linebreak

\begin{figure}
\begin{center}
\leavevmode \epsfxsize=4.5 cm  \epsfbox{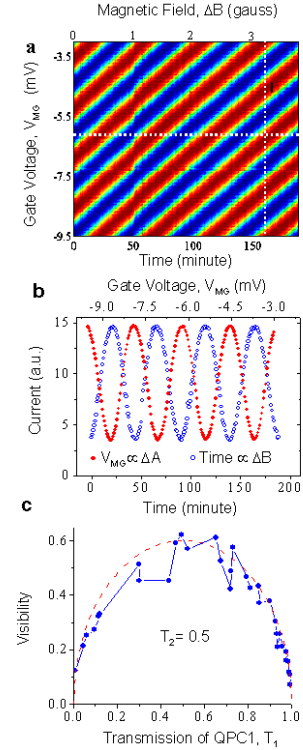}
\end{center}
\vspace{0 cm} \caption{Interference pattern of electrons in a Mach
Zehnder interferometer and the dependence on transmission. (a) Two
dimensional color plot of the current collected by \textbf{D1} as
function of magnetic field and gate voltage at an electron
temperature of $\sim$ 20mK.  The magnet was set in its persistent
current mode ($B \sim$ 5.5T at filling factor 1 in the bulk) with
a decay rate of some 0.12 mT/hour, hence time appears on the
abscissa. The two QPCs were both set to transmission
$T_1=T_2=0.5$. Red (blue) stands for high (low) current. (b) The
current collected by \textbf{D1} plotted as function of the
voltage on a modulation gate (red plot) and as function of the
magnetic field (blue plot) - along the cuts shown in a.  The
visibility of the interference is 0.62.  (c) The visibility of the
interference pattern as a function of the transmission probability
$T_1$ of \textbf{QPC1} when \textbf{QPC2} is set to $T_2=0.5$.
Dashed line is a fit to the experimental data with
visibility=2$\eta \sqrt{T_1(1-T_1)}$.  The normalization
coefficient $\eta$=0.6 accounts for possible decoherence and/or
phase averaging.}
\end{figure}

\noindent
We first test the ideality of the Ohmic contacts and the
validity of the edge states picture.  For both QPCs open a nearly
ideal Hall plateau was observed in $I_{D1}$ while no current was
measured in \textbf{D2} ($I_{D2}$=0).  That validated that current
was confined to the outer edge with no backscattering across the
3$\mu$m wide mesa. We then pinched off QPC1 or QPC2 and found
again a Hall plateau in $I_{D2}$ with zero current in \textbf{D1}
($I_{D1}=0$). This proved that the small Ohmic contact of
\textbf{D2} was ideal and fully absorbed the current. Setting then
both QPCs to $T_1=T_2 \sim 1/2$ and varying the magnetic field $B$
(actually the time) or the area $A$ (the voltage on a \textbf{MG})
lead to pronounced interference signal in \textbf{D1} (or in
\textbf{D2}) with visibility as high as 0.62 (Fig. 2). Since the
field decays linearly with time and the area (or electron density)
vary proportional to the gate voltage changing these parameters
leads to the diagonal straight color lines (of constant phase)
seen in Fig. 2a.  Figure 2b shows similar data taken along two
cuts (the dotted lines shown in Fig. 2a) - one for constant $B$
and one for constant $A$.  The cleanliness of the interference
pattern and the high visibility prove the nearly ideal nature of
the interferometer.

In order to further verify the \textit{two-path nature} of the
interference the visibility was measured as function of $T_1$ for
a constant $T_2=0.5$ (see Fig. 2c).  It agrees well with the
expected expression for the visibility
$\upsilon=2\eta\sqrt{T_1(1-T_1)}$, with $\eta \sim 0.6$ a
normalization factor that accounts for dephasing (either due to
phase averaging in the energy window of the electrons or due to
inelastic scattering processes). Moreover, the period of the
oscillations, in time and in \textbf{MG} voltage, agrees well with
one flux quantum being added (or subtracted) in the ring's area.
The time period is $\sim$50min, which is the time needed for one
flux quantum decay in the superconducting magnet, while the
voltage period agrees approximately with that needed to deplete
one electron (hence, one flux quantum for filling factor 1) under
the \textbf{MG} gate.

While the visibility is very high it is still smaller than unity
($\upsilon \sim0.6$).  An obvious reason is the finite energy
spread of the electrons at the edge (due to their finite
temperature) and the unavoidable dependence of the AB area on the
energy (hence, the AB phase) - leading to phase averaging
(\textit{thermal smearing}). Indeed, the visibility was found to
drop precipitously with increasing temperature or applied voltage
at \textbf{S}, as seen in Fig. 3. In this example, a mere increase
of the temperature to 100mK (some $9\mu$eV) reduced the visibility
from $\upsilon \sim 0.53$ to $\upsilon \sim 0.01$ (plotted in red
in Fig. 3a).  If indeed phase averaging is the cause for the
dephasing, it could, in principle, be eliminated with
monoenergetic electrons.  A minute AC signal ($\sim0.5\mu$V) at
1.4MHz was added to a variable DC voltage $V_{DC}$ and the
synchronous AC part of the interfering signal was measured at
20mK.  This signal leads to a \textit{differential visibility}
$\upsilon_d$, resulting only from the electrons in an energy
window $\sim0.5\mu$eV around an energy $eV_{DC}$. Surprisingly, as
seen in Fig. 3a (plotted in blue), the energy dependent
differential visibility at $T$=20mK is strikingly similar to the
temperature dependent visibility with a relation between the
scales $eV_{DC} \sim 4k_BT$. The visibility (in color scale) is
plotted as function of both $T$ and $V_{DC}$ in Fig. 3b.  The
clear symmetry across the diagonal suggests that the dephasing
processes due to temperature and voltage are similar.
Unfortunately this contradicts our previous assertion of phase
averaging taking place in a wide window of energy and points at
decoherence, induced by inelastic scattering events, as the main
source of dephasing.  In other words, for an increased temperature
or for high energy monoenergetic electrons, empty states are being
created allowing energy loss via scattering. \linebreak

\begin{figure}
\begin{center}
\leavevmode \epsfxsize=5.5 cm  \epsfbox{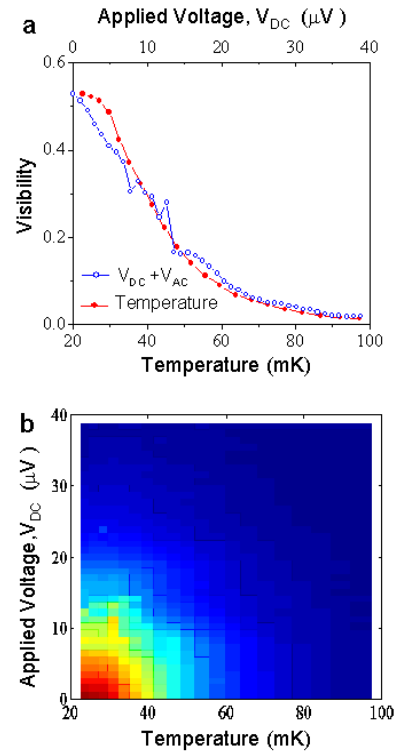}
\end{center}
\vspace{0 cm} \caption{The dependence of the visibility of the
interference pattern on temperature and applied voltage. (a)
Visibility as function of temperature at small excitation voltage
for $V_{DC}=0$ (red plot), and as function of $V_{DC}$ with a
small AC voltage $V_{AC}$ superimposed on it at electron
temperature 20mK (blue plot).  Both QPCs were set to
$T_1=T_2=0.5$. (b) A 2D color plot of the visibility as function
of temperature and applied DC voltage. Red (blue) stands for high
(low) visibility. }
\end{figure}

In order to test this hypothesis current shot noise was measured.
Its spectral density, defined as the averaged square of the
current fluctuations per unit of frequency, $S=\langle (i^2)
\rangle /\Delta f$, and for stochastic partitioning at zero
temperature $S \propto eV_{DC}T_{SD}(1-T_{SD})$~\cite{r13}.
Introducing a phenomenological parameter $k$ that accounts for
decoherence in the interferometer with $T_1=1/2$ and
$T_{SD}=0.5+k\sqrt{T_2(1-T_2)cos\varphi}$, we find that for
complete phase averaging or for a complete decoherence $T_{SD}=0$.
On the other hand shot noise in \textbf{D1} is $S_{D1}\propto
T_{SD1}(1-T_{SD1})=1/4-k^2T_2(1-T_2)cos^2\varphi$, with
$S_{D1}$=const. for $k=0$ but $S_{D1}=1/4-k^2T_2(1-T_2)/2$ for
complete phase averaging (resulting from an integration of
$cos^2\varphi$ in the range $\varphi=0....2\pi$). Hence, noise is
expected to exhibit a parabolic dependence on $T_2$ in a coherent
system. Shot noise was measured (see Refs. 12 and 13 for details)
with a relatively large $V_{DC}$ applied at \textbf{S} so that
interference signal was quenched (negligible visibility).  The
dependence of $S$ on $T_2$, shown in Fig. 4, followed the above
expression with $k \sim 0.9$, proving that indeed phase averaging
is dominant while decoherence is negligibly small.

\begin{figure}
\begin{center}
\leavevmode \epsfxsize=8.5 cm  \epsfbox{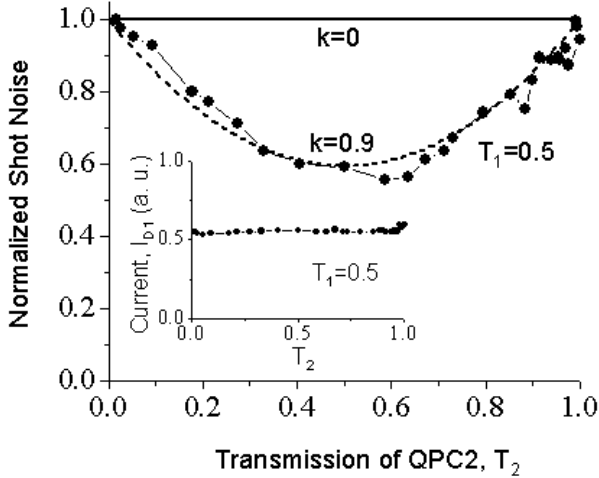}
\end{center}
\vspace{0 cm} \caption{Shot noise measurement (at filling factor
2) as function of $T_2$ when the transmission of QPC1 was set to
$T_1=0.5$. A 30$\mu$V DC voltage (under which the AB interference
pattern was quenched) was used to measure shot noise.  The shot
noise of the current collected by \textbf{D1} is shown by the
black dots (normalized to a maximum), while the two solid lines
are the expected noise for $k=0$ and $k=0.9$, respectively,
according to the simple model described in the text.  The
agreement with the simple model indicates that the electrons are
coherent even at the DC voltage where the interference pattern
fades away.  Inset: the current at $D_1$ as function of $T_2$ at
$V_{DC}=30\mu$V.  As expected from the lack of the interference
pattern, the current is independent of $T_2$ (see text). }
\end{figure}

A single particle, namely, a non-interacting model would lead to
the following dependences of the visibility on energy: for $V=0$
and finite $T$, $\upsilon \propto \beta T/sinh(\beta T)$, with
$\beta$ a constant; for finite $V$ but $T=0$, $\upsilon \propto
sin[(e/2\pi)V]/[(e/2\pi)V]$, while the differential visibility at
$T=0$ is expected to be voltage independent.  Since the
experimental results contradict these projections we propose (with
no proof yet) two possible reasons for the dephasing.  One might
be due to low frequency noise (say, $1/f$ type due to moving
impurities), which might be induced by a higher current, leading
to fluctuation in the area and consequently, phase smearing.  The
other could be related to the self consistent potential contour at
the edge. Since it depends on the local density of the electrons
in the edge state~\cite{r14}, fluctuation in the density due to
partitioning are expected to lead to fluctuation in the AB area
enclosed by the two paths and hence to phase randomization.  For
example, for $B \sim$ 5.5T a merely 1$\sim$2 angstroms shift of
the edge suffices to add one flux quantum into the enclosed area.

Our aim here was to present a novel and powerful electron
interferometer, which might to be used as a powerful tool for
future interferometry studies of electrons.  One exciting
possibility is the study of coherence and phase of fractionally
charged quasiparticles in the fractional quantum Hall effect
regime~\cite{r15}.

\textit{Acknowledgements} We thank Y. Levinson for clarifying the
issue of phase averaging and C. Kane for providing useful comments
on the manuscript.  The work was partly supported by the MINERVA
Foundation, the Israeli Academy of Science, the German Israeli
Project Cooperation (DIP), the German Israeli Foundation (GIF),
and the EU QUACS network.

\end{multicols}
\end{document}